# Probing many-body localization in a disordered quantum magnet

D.M. Silevitch[1], C. Tang[1], G. Aeppli[2], T.F. Rosenbaum[1]

[1] Division of Physics, Mathematics, and Astronomy, California Institute of Technology, Pasadena, California 91125, USA

[2] Laboratory for Solid State Physics, ETH Zurich, Zurich, CH-8093, Switzerland; Dept. de Physique, EPF Lausanne, Lausanne, CH-1015, Switzerland; and Swiss Light Source, Paul Scherrer Institut, Villigen PSI, CH-5232, Switzerland

## ABSTRACT

Quantum states cohere and interfere. Quantum systems composed of many atoms arranged imperfectly rarely display these properties. Here we demonstrate an exception in a disordered quantum magnet that divides itself into nearly isolated subsystems. We probe these coherent clusters of spins by driving the system beyond its linear response regime at a single frequency and measuring the resulting "hole" in the overall linear spectral response. The Fano shape of the hole encodes the incoherent lifetime as well as coherent mixing of the localized excitations. For the disordered Ising magnet, $LiHo_{0.045}Y_{0.955}F_4$, the quality factor $Q$ for spectral holes can be as high as 100,000. We tune the dynamics of the quantum degrees of freedom by sweeping the Fano mixing parameter $q$ through zero via the amplitude of the ac pump as well as a static external transverse field. The zero-crossing of $q$ is associated with a dissipationless response at the drive frequency, implying that the off-diagonal matrix element for the two-level system also undergoes a zero-crossing. The identification of localized two-level systems in a dense and disordered dipolar-coupled spin system represents a solid state implementation of many-body localization, pushing the search forward for qubits emerging from strongly-interacting, disordered, many-body systems.

Localization in quantum systems remains both fundamental to science as well as to technology. It is a venerable subject, starting with the work of Anderson[1]—whose name is associated with disorder-induced localization—and Mott, whose Mott localization transition[2] is due to repulsion between electrons. Since these pioneering studies, similar concepts have been extended beyond correlated electrons to a wide range of interacting quantum systems. The combined problem of many-body localization[3-10] persists to this day, and has inspired recent numerical simulations[11,12] as well as experiments on cold atoms[13-15]. There is also practical relevance for systems of quantum devices: a notable example is the "D-wave" processor[16,17], which attempts to implement adiabatic quantum computation, but may be limited as a matter of principle by localization effects.

To control these long-lived and independent states, it is necessary to know how they interact with each other and with the outside world. Minimizing the interactions between coherent localized states and the continuum of states in the broader environment is an important goal for realizing an effective quantum computer[18-21]. However, these environmental couplings are by definition weak compared to the transitions among the states contributing to the spectrum for a particular localizing environment, making it difficult to study them directly. Recently, it has been posited that many weak couplings of this sort can be probed by pumping the system into a nonlinear response regime[22], saturating the discrete transition associated with the coherent state, and resulting in spectral holes. In the present work, we use the Fano lineshapes of the spectral holes to characterize the coupling between the localized subsystems and both external and internal fields, and so establish $LiHo_xY_{1-x}F_4$ as a solid state system exhibiting many-body localization. Our work differs from experiments in quantum optics[23] in that we are examining emergent degrees of freedom in a (magnetic) many-body system rather than single-particle states in atoms and semiconductor quantum dots. At the same time, it represents a major advance over our own previous activity on pump-induced Fano resonances in the same magnetic material[24,25] in that we uncover a remarkably simple phenomenology, including the discovery of a zero-crossing for the Fano asymmetry parameter $q$, as a function of non-linear drive amplitude and quantum mixing via a transverse field.

Asymmetric absorption lineshapes in atomic gases were first addressed by Ugo Fano more than 50 years ago[26]. They arise from interference between discrete transitions in the atoms and ionization into the continuum. Now known as Fano resonances, this formalism has found wide applicability in systems ranging from photo-ionized gases[27] to high-$T_c$ superconductors[28] to photonic crystals[29] to quantum wells[30]. Their extension to spin liquids provides a means to characterize coherent spin clusters labeled in the time/frequency domain but distributed randomly in space.

The remainder of the paper is structured as follows. We begin by presenting an overview of the experimental system, the dilute Ising magnet LiHo$_x$Y$_{1-x}$F$_4$, focusing on the behavior at small $x$. We then develop a framework for understanding the linear and nonlinear susceptibilities in terms of individual spins and coherently-bound clusters of these spins. In the Results section, we describe a pump-probe technique to study these clusters, using the detailed lineshape of the observed Fano resonances to track the evolution of the clusters as a function of external variables such as temperature and applied magnetic field. Most notably, we find that the interaction between the clusters and the incoherent bath of free spins can be tuned continuously to a point where the direct absorption of energy, and hence dissipation, of the clusters goes to zero. This dissipationless response suggests that the cluster behavior can be understood in the framework of many-body localization effects. Finally, in the Discussion section, we make explicit the connection of our experiments to a common (spin-based) realization of many-body localization, as well as derive the meaning of the zero crossing of the Fano asymmetry parameter $q$ in terms of the relevant matrix elements.

### Review of experimental system $LiHo_xY_{1-x}F_4$

Spin clusters functioning as quantum two-level systems form in the dilute, dipolar-coupled magnet $LiHo_xY_{1-x}F_4$ under the appropriate thermodynamic conditions[25,31]. The magnetism in this family of rare-earth fluorides has long been studied as a realization of the dipole-coupled S=1/2 Ising model, with the spins carried by the $Ho^{3+}$ ions and the Ising axis lying along the crystallographic $c$ axis[32,33]. The non-magnetic $Y^{3+}$ ions randomly occupy the same sites as the magnetic $Ho^{3+}$ ions with probability 1-$x$. The hierarchy of quantum levels accounting for the charge neutral excitations of individual $Ho^{3+}$ ions (in the dilute limit where $x<<1$) of this wide-gap insulator has been summarized recently by Matmon et al.[34] Most relevant for the current low temperature study are the ground-state doublet for the Ising spins with a crystal field-derived 9.4 K gap to the first excited state and the hyperfine interaction between the electronic ($J=8$) and nuclear ($I=7/2$) spins of the $Ho^{3+}$ ions[32] that yields a nuclear Zeeman ladder consisting of eight states with spacing 0.2 K between consecutive levels. Analytic solutions of the microscopic Hamiltonian,[35-38] combined with measurements of the crystal-field parameters,[32,39] have quantitatively connected the microscopic Hamiltonian to the long-standing effective Hamiltonian for the spin physics. Magnetic fields applied perpendicular to the Ising axis mix the ground state doublet with the first excited state, inducing a splitting of the doublet which in the low field limit scales as $\Gamma \propto H_t^2$ (in contrast to the Zeeman splitting $\propto H_l$ of the ground state in a longitudinal field $H_l$) and leads to an effective transverse field Ising Hamiltonian[40,41]:

$$H = \sum_{i,j} J_{ij} \sigma_i^z \sigma_j^z - \Gamma \sum_i \sigma_i^x \,. \qquad (1)$$

In the pure ($x=1$) limit, a classical ferromagnetic transition occurs at the Curie point $T_C=1.53$ K dictated by the magnitude of the (predominantly) dipolar interactions $J_{ij}$. The quantum fluctuations induced by the transverse field disorder the Ising spins, producing a quantum critical point at $\Gamma = 1.6$ K where the Ising ferromagnetism vanishes. This zero temperature transition is linked by a line of second order transitions between paramagnetic and ferromagnetic phases to zero field. The principal features of the dynamics are propagating soft magnetic modes[42].

While Eq. (1), which takes no account of the nuclear spins, is an excellent starting point for understanding the physics of pure LiHoF$_4$, at temperatures below ~0.6 K the electronic and nuclear spins of the Ho$^{3+}$ ions[32] combine to form composite degrees of freedom with effective spins $I+J$, resulting in an upturn in the ferromagnet-paramagnet phase boundary for pure LiHoF$_4$[41]. Furthermore, entanglement of the nuclear and electron spins results in an incomplete softening of the principal magnetic excitation mode at the quantum phase transition[39,42].

Additional physics has been revealed at holmium concentrations between $x=1$ for the pure ferromagnet and the $x\ll1$ dilute ion limit[43]. The combination of disorder, the magnetic dipole interaction, which can be ferromagnetic or antiferromagnetic (depending on the relative orientation of the preferred spin direction and the vector separating two spins), and quantum fluctuations[44-46] creates a sequence of states as a function of decreasing dipole (Ho) concentration[43], progressing from mean-field ferromagnet[41] to random-field ferromagnet[47] to spin glass[40,48-50] to spin liquid[31]. The quantum fluctuations, which arise from internal transverse fields generated by the loss of translational symmetry as a result of dilution, act to prevent complete freezing in the $T \rightarrow 0$ limit even without an externally-applied transverse field[51]. Here, we focus on the Bhatt-Lee spin-liquid[52] (originally proposed for phosphorus-doped silicon near its metal-insulator transition) for x ~ 0.05 and weak thermal coupling, characterized by a hierarchy of singlets derived from combinations of doublets (isolated spins) and triplets (ferromagnetically coupled spins). The hierarchy of singlets results in a low frequency susceptibility which scales as $1/T^\alpha$ with $\alpha \neq 1$ rather than following Curie or spin-glass forms. For LiHo$_{0.045}$Y$_{0.955}$F$_4$, $\alpha$ was experimentally measured to be 0.75, less than the Curie exponent $\alpha = 1.0$ due to the compensation of spins by each other as they form singlets on cooling.[51] The nature of the ground state for LiHo$_{0.045}$Y$_{0.955}$F$_4$ has been a matter of debate, with some groups reporting a spin-glass state[49,53] and other work indicating a spin-liquid state[31,43]. This discrepancy was resolved in Ref. 25, which showed that by varying the strength of thermal coupling between the crystal and an external heat bath, it was possible to tune between the two limits. This controlled the rate at which lattice phonons, and subsequently spins via spin-lattice coupling, could exchange energy with the external bath. As discussed in the Results section below, the measurements reported here were all obtained in the weak-thermal-coupling limit which favors the quantum spin liquid.

Fig. 1a shows schematically how a disordered quantum spin system such as LiHo$_x$Y$_{1-x}$F$_4$ at low concentrations or Si:P[52,54] breaks into decoupled clusters and isolated spins, focusing on some clusters, whose classical ground states are ferromagnetic but which because of small transverse fields imposed by other clusters, can be described as two-level systems with low-energy eigenstates which are coherent superpositions of up and down configurations $|\Uparrow> \pm |\Downarrow> = |\uparrow\uparrow \cdots \uparrow\rangle \pm |\downarrow\downarrow ... \downarrow\rangle$.[31]

More elaborate wavefunctions and level schemes[42,55], taking account of e.g. electronuclear spin mixing and also classical ground states with antiferromagnetic correlations, will not qualitatively change the physics of hole burning which we seek to outline in this section, but will be invoked later when we discuss certain details of our data. The key point is that when the clusters are independent, the effective low-energy Hamiltonian is $H_{le} = \sum_i H_i$ with

$$H_i = \Delta_i \sigma_i^x + M_i h(t)\sigma_i^z , \qquad (2)$$

where the sum is over decoupled clusters $i$ characterized by an underlying moment $M_i$ and pseudospin operators $\sigma_i$, $\Delta_i$ corresponds to the splitting between the two states $|\Uparrow\rangle \pm |\Downarrow\rangle$ for cluster $i$, and $h(t)$ is an external drive field. If $h(t)$ is a time-independent constant $h>0$, the magnetization normalized to $M_i$ is

$$\langle \sigma_i^z \rangle = \frac{1}{2}\frac{M_i h}{\lambda_i} , \qquad (3)$$

where $\lambda_i = \sqrt{\Delta_i^2 + h^2 M_i^2}$. For $hM_i \ll \Delta_i$, $\langle \sigma_i^z \rangle = \frac{1}{2} M_i h/\Delta_i$ while for $hM_i \gg \Delta_i$, we obtain the expected saturation value of 1/2. Now we consider an oscillating field $h(t) = h\cos(\omega t)$ where for all $i$, $\hbar\omega \ll \lambda_i$. The full magnetization in this case will then merely be $M(t) = \sum_i M_i <\sigma_i^z> \cos(\omega t)$. If we relax the condition on $\hbar\omega$ but insist that the drive amplitude $h \ll \Delta/M_i$, we obtain the usual linear response form

$$M(t) = \sum_i M_i^2 h \left\{ \left(\frac{\Delta_i}{\Delta_i^2 - (\hbar\omega)^2}\cos(\omega t)\right) + \sin(\omega t)\big(\delta(\hbar\omega - \Delta_i) - \delta(\hbar\omega + \Delta_i)\big) \right\} \qquad (4)$$

where there is now an out-of-phase response which if the frequency is scanned gives the density of states, weighted by $M_i^2$ for the clusters.

Once we see a continuum in the out-of-phase linear response for a many-body system, a priori we do not know whether we are dealing with a sum of spectra of localized subsystems, as in Eq. (2), for which the eigenfunctions are simply given by direct products of the wavefunctions for the subsystems, or if we are dealing with shorter-lived excitations whose behavior is dominated by coupling between subsystems. A standard method in spectroscopy to determine whether a continuum is due to independent – i.e. localized – two-level systems is to simultaneously relax the conditions on $\omega$ and $h$ to enter the more interesting and heavily studied regime of driven two-level systems. The idea is to apply a large amplitude field at frequency $\hbar\omega = \Delta_i$ so that $\langle\sigma_i^z(t)\rangle$ has oscillations of sufficiently large amplitude so that further increments in field can yield only small increases in $\langle\sigma_i^z\rangle$. Without delving into the mathematics of the time-dependent Schrödinger equation[56] for the Hamiltonian (1), one can convince oneself (a) of the plausibility of this given the magnetization saturation with increasing $h$ in the static limit described by Eq. (3). At the same time, (b) the resonant enhancement of the linear response (4) for $\omega$ near $\Delta_i/\hbar$ indicates that we can preferentially excite subsystems $i$ with $\omega = \Delta_i$. (a) and (b) together lead to the conclusion that the sample becomes more transparent to radiation at the drive frequency, and a sharp hole is burnt into the spectrum if the two-level system in question cannot interact with two level systems with other values of $\Delta_i$. If such spectral holes can be found in an interacting many-body system, then we can speak about the system exhibiting many-body localization in the sense of having excited states which are direct products of the excited states for subsystems, meaning that the entire system cannot act as its own heat bath. If there is a weak residual coupling between subsystems, the holes will acquire shapes with width parameters which measure that weak coupling.

**Results**

For LiHo$_{0.045}$Y$_{0.955}$F$_4$ both the thermal link to the heat bath[25] and magnetic fields applied transverse to the Ising axis[24] control the observability of localized excited states. In particular, strong coupling to a thermal bath yields the linear response expected from a spin-glass[25,49], no observable localized excited states, and consequently no evidence for the Fano resonance is seen for strong coupling (Fig. 1d). We therefore concentrate here on weak coupling to the bath.

The first step in the measurements is to determine pump fields $h_{pump}$ sufficient to enter the non-linear regime. Fig. 1c shows the magnitude and phase angle of the magnetization induced for a range of frequencies $f_{pump}$, limited to a maximum of 10 Hz to allow measurement to drive fields above 1 Oe without excessive eddy current heating. The data generally follow an S-shaped curve, whose inflection point moves outward with increasing frequency, in agreement with the model (2) where the longitudinal field couples the localized singlet ground state of a cluster to an excited state separated by a small gap[51]; the susceptibility at higher frequencies is more sensitive to smaller clusters where the gap is larger, which can only be overcome by a higher drive field $h_{pump}$. For our pump-probe experiments we operate near to and somewhat above the inflection point; very high driving fields are avoided because they lead to excessive heating. The effective increase in energy of the spin system can be characterized by measuring the shift in the peak frequency of $\chi''(f)$ as a result of nonlinear drive and comparing with linear response measurements at a range of temperatures. As reported in Ref. 31, a 5 Hz, 0.2 Oe pump provided energy equivalent to a 40 mK increase in temperature. This can be distinguished from purely thermal effects associated with the drive field by noting the clear change in the dissipative lineshape: the distribution of oscillators visible in $\chi''(f)$ lacks the tails seen at equilibrium (in the absence of pumping). While the system consists of an ensemble of spin clusters with a broad range of sizes (previous magnetization measurements found clusters ~250 spins at a 5 Hz resonant frequency), choosing a particular pump frequency selects for a set of clusters which are similarly sized and hence sorted by resonant frequency. These clusters are largely independent of each other, and hence pumping at two well-separated frequencies simultaneously results in two distinct holes in the overall spectrum.[31]

We turn now to pump-probe measurements, examples of which are shown in Fig. 2a. In these measurements, the system is driven with a strong ac pump field at frequency $f_0$ and probed with a weak field at a range of frequencies $\Delta f = f_{probe} - f_0$ around $f_0$ (see Methods for details). The characteristic asymmetric shape of a Fano resonance is immediately obvious, providing a direct indication of a weak coupling between the coherent spin cluster and the continuum of surrounding spin states (Fig. 1b,c). By comparing the integrated area of the resonant response with the area of the entire linear-response spectrum, we estimate that the fraction of spins bound in clusters resonant at the chosen drive frequency is of order $2 \times 10^{-6}$ of the entire sample. As the temperature is increased, the amplitude of the resonant response drops, and the resonance appears to broaden, with the response suppressed to 8% of its original amplitude at $T = 500$ mK and to below the noise floor of the measurement at 700 mK. Given that the overall linear susceptibility of LiHo$_{0.045}$Y$_{0.955}$F$_4$ has a strong temperature dependence, the thermal evolution of the resonant response can be seen more clearly by normalizing it to the linear response at each temperature (determined by measuring $\chi''(\Delta f = 30$ mHz$)$). We show in Fig. 2b spectra obtained at a series of temperatures, normalized, and then combined into a surface plot where color and height now represent the absorption for a given $\Delta f$ and $T$. The broadening of the resonance with increasing temperature emerges clearly in this visualization, and we examine it quantitatively in Fig. 2c by looking at the evolution of the linewidth in the fits to the Fano form,

$$\chi''(\Delta f) = A \frac{\left(\frac{q\Gamma}{2} + \Delta f\right)^2}{\Delta f^2 + \frac{\Gamma^2}{2}}, \qquad (5)$$

where $\Gamma$ is the linewidth of the resonance and $q$, known as the Fano parameter, characterizes the interference between the different transition pathways. The mHz scale low-temperature limit of the linewidth suggests that the coupling between the clusters and the background spin bath is weak, and hence that the system can be considered in the framework of a 2-level system in weak contact with the environment rather than a continuous relaxation process. Quantum states with splittings substantially smaller than nominal bath temperatures are very common in solids and liquids, and indeed form the basis for various resonance (e.g. NMR) spectroscopies, many of which rely on non-equilibrium quantum state preparation. Reduced bath coupling during cool-down increases the $T_1$ and $T_2$ times associated with such quantum states, and so makes a description of the magnetic response of the system as due to a set of independent multilevel quantum systems more appropriate than a picture based on classical, thermal diffusion.

When a multi-level quantum description applies, for a fixed bath coupling (which extracts heat) and ac driving field (which inserts heat), an equilibrium with a set of fixed state occupancies will characterize the system, and to first order the equilibrium can be described by a fixed effective temperature. On the other hand, the non-equilibrium dynamics are dominated by small multilevel systems describable in terms of some generalized Bloch equation, exactly as is the case e.g. for NMR performed even at room temperature, and are therefore quantum mechanical. The linewidth increases exponentially with $T$, consistent with a thermally activated process with a gap $\Delta = 740$ mK, which is an energy well below the 9.4 K first excited crystal field state energy but of the same order as nearest neighbor spin couplings as well as the energy difference[34,41,57] (~750 mK) between electronuclear states with nuclear moments of 7/2 and ½. This suggests that hole burning is favored when Ho nuclear spins are not relaxing relative to electron spins, but does not exclude composite electro-nuclear wavefunctions involving several Ho ions. Indeed, that several ions need to be involved can be deduced from comparison of the crossover fields (all below 1 Oe) in our ac drive data (Fig. 1c) with the step locations at multiples of 200 Oe in the magnetometry of Giraud et al.[55] The greatly increased density of level crossings suggested by our data follows naturally from the dipolar interactions between multiple Ho ions. Further indications that the interactions are important come from measurements where changing the thermal boundary conditions vary the state of the system[25]. For strong coupling to an external bath (the mixing chamber of the cryostat), collective spin glass behavior obtains, and the nonlinear response also changes qualitatively, with no trend towards saturation. In other words, the resonances are not due to single-spin behavior but instead are consistent with a picture consisting of clusters of many spins behaving as a large effective 2-level system protected from others in a wavefunction such as that $|\Uparrow> \pm |\Downarrow> = |\uparrow\uparrow \cdots \uparrow\rangle \pm |\downarrow\downarrow \ldots \downarrow\rangle$ (discussed above) at low temperatures because of a relatively large single spin flip energy, associated with either the hyperfine or electronic dipole interactions. In addition to the thermal broadening of the resonance, the lineshape asymmetry *q(T)* is also *T*-dependent, varying in approximately linear fashion for *T* < 0.25 K, and plateauing above that point (Fig. 2d). A possible microscopic origin for this behavior is that as the temperature grows, thermally activated spin flips will occur within clusters containing antiferromagnetically coupled spins. Such spin flips will result in increased dipolar moments for the clusters, and thus an increased coupling to remnant dipole moments of the other clusters forming the underlying "bath", an effect seen in the temperature-dependent linewidth shown in Fig 2c. To rephrase, heating begins to unbind the clusters and so reveals them to each other. As this occurs, there will also be loss of many-body localization, and the population of localized subsystems that can still interact coherently with the driven subsystem will drop, implying an increase in the residual response at the drive frequency which, according to Eq. (5) evaluated at *Δf=0,* translates directly into an increased *q*. Provided that a single constant $J_{AFM}$ characterizes the underlying antiferromagnetic coupling, we would find that for $k_BT \gg J_{AFM}$ the probabilities that two spins are either ferromagnetically or antiferromagnetically correlated become equal and we would see a plateau in the coupling to other clusters as well as the value $q$. Such plateaus also can be found if there is a discrete series of antiferromagnetic couplings $J_{AFM,i}$ and the conditions $J_{AFM,i} \ll k_BT \ll J_{AFM,i+1}$ are met. The discrete nature of the distribution of dipolar couplings for the $LiHo_xY_{1-x}F_4$ lattice leads to the possibility that these conditions obtain, and therefore the data in Fig. 2d, which extend only as far as hole burning can be seen (and $q$ can be measured), could be a manifestation of such a plateau.

It should be noted that the "free-induction" relaxation time of ~10-30 seconds observed previously[31] is substantially shorter than the ~500-1000 seconds of the inverse linewidth of the hole uncovered in the driven pump-probe measurements. This follows because the "free induction decay" was measured for relaxation after the strong ac drive field was turned off, while the linewidth here is measured in the far more weakly driven linear regime. More formally, the rotating wave approximation[56] does not apply for the combination of strong (non-linear) drive fields and low frequency employed for our experiments. In particular, the Rabi frequency $f_{Rabi}$ associated with the 0.5 Oe drive field $h_{pump}$ for the electronic (Ising-like) spin of a single $Ho^{3+}$ ion is $g_\parallel \mu_B h_{pump} \sim 10$ MHz $\gg f_{pump}$, which is precisely opposite to the requirement that $f_{Rabi} \ll f_{pump}$ for the rotating wave approximation to hold.

We explore in Figs. 3 and 5 the effects of changing the amplitude of the pumping field. Most important is the change in sign of the Fano $q$: for the largest drive field (0.5 Oe) the low and high frequency responses are enhanced and suppressed, respectively, opposite to what we see for the lower drive field. The zero crossing of $q$ occurs at a critical $h_{pump} = 0.45$ Oe (Fig. 5a). The data points at the pump frequency at which pump and probe-derived signals cannot be distinguished are ignored for the Fano fits because they represent the response of the highly excited (pumped) clusters and not the perturbatively mixed clusters with other resonant frequencies.

Panels (b) and (c) of Fig. 5 reveal clear distinctions between $\chi_{drive}$, the total signal at $f_{pump}$ and $\chi_{Fano}$, the linear Fano contribution calculated from evaluation at $f = f_{pump}$ of the fitted Fano form to data at $f \neq f_{pump}$. First, $\chi'_{drive}$ goes through a maximum at the zero-crossing of $q$ (Fig. 5(c)), while $\chi'_{Fano}$ undergoes a decrease which looks like a rounded step. Second, when we plot the phases $\phi = \tan^{-1} \chi''/\chi'$ (Fig. 5(b)), we find that while both $\chi_{drive}$ and $\chi_{Fano}$ have phase shifts which are smaller at high $H_{drive}$, the latter actually has a zero near the zero of $q$. In other words, for small linear perturbations, the Fano response is dissipationless in the limit $f \to f_{pump}$. This result follows from Eq. 5, which gives $\phi = \phi(q) = \tan^{-1} \frac{q^2}{1-q^2}$, a functional form that we superpose over the experimental results in Fig. 5(b). The absence of dissipation in the Fano response that describes the linear continuum at $q=0$ means that hole burning is actually complete at the drive frequency: there is no continuum contribution to $\chi''$ which remains unaffected by the drive in the limit $f \to f_{pump}$. Significantly, the absence of dissipation coincident with the $q$ zero crossing indicates that the clusters cannot be excited between their ground and first excited states by the external drive. While $\phi(q)$ gives a rough account of the experimental phase angle as $q$ moves away from zero, the data ultimately deviate from $\phi(q)$, which implies that some oscillators even with frequencies close to $f_{pump}$ are not contributing to the Fano profile.

Varying the ac pump amplitude accesses different mixtures of the states of the localized clusters. The additional power applied to the drive solenoid also results in eddy-current induced heating of the copper susceptometer mount, and hence some degree of conductive heating of the sample despite the low-thermal-conductivity Hysol epoxy spacers holding the sample inside the susceptometer. This, as well as dissipation within the sample itself, gives rise to a higher effective temperature, with a concomitant loss of coherence. The decoherence of the resonant excitation is reflected by a measurable increase in the linewidth $\Gamma_r$, whose temperature-dependent evolution can be traced readily in Fig. 2c. Over the range of pump amplitudes shown in Fig. 3, the resonance linewidth increases from 1.1 mHz to 1.8 mHz (Fig. 5d), equivalent to approximately 50 mK of direct thermal heating. Even while heating is clearly present, the line widths remain negligible on the scale of the drive frequency, allowing the coherent superpositions of the cluster and continuum oscillations. Their relative signs change at a critical longitudinal pump field of 0.45 Oe, thus yielding the zero crossing of $q$, one of the main results of our experiment.

We now take advantage of one of the key features of the Li(Ho,Y)F$_4$ family, the ability to tune the microscopic Hamiltonian by applying a magnetic field transverse to the Ising axis, thereby quantum-mechanically mixing the single ion and cluster eigenstates[35,40] via different matrix elements than does the ac longitudinal field. Nonetheless, application of a transverse field induces a crossover (Fig. 4) at a well-defined field of $H_t$ = 3.5 kOe[24], similar to that seen earlier as a function of pump amplitude (Fig. 3). We plot in Fig. 6(a) the transverse field dependence of the Fano parameter $q$. This parameter changes linearly with $H_t$ over most of the experimental range, showing that the external transverse field not only changes the energies of different states, but also tunes their interactions with the broader spin bath environment. In particular, at the $H_t$ =$H_c$=3.5 kOe crossover field, $q$ vanishes. At the same time, as also seen when we varied $h_{pump}$ to obtain a zero-crossing of $q$, there is a quadratic zero in the phase for $\chi_{Fano}(f_{pump})$ (Fig. 6b) and a maximum in $\chi'_{drive}$ (Fig. 6c), both of which coincide with the zero of $q$ and can be roughly described by the function $\phi(q)$. The simultaneous vanishing of $q$ and the phase (and hence the dissipation) going to zero opens the possibility for the static transverse field to be used to decouple the localized two-level systems from the external ac field. In contrast to what we saw for the $h_{\text{pump}}$ scan with $H_t=0$, $\chi'_{\text{drive}}$ and $\chi'_{\text{Fano}}$ are nearly indistinguishable at $f_{\text{pump}}$. Another contrast, anticipated from the previous paragraph and visible in the comparison of Figs. 5d and 6d, is that the linewidth is, to within error, $H_t$ – independent. The essentially constant behavior of the linewidth is a strong indication that the evolution due to the transverse-field-induced quantum fluctuations is fundamentally different from the purely classical behavior seen as a function of increasing temperature.

**Discussion**

Even if only a subset of the spin clusters in driven $LiHo_{0.045}Y_{0.955}F_4$ display localized excited states, the material with its relatively high density of interacting spins is a promising venue to investigate many-body localization (MBL). The discrete two-level spin clusters are formed of finite numbers of spins that are locked in orientation relative to each other, with a renormalized dipolar coupling to the other spin clusters. As the data have shown, when driven strongly with a 'pump' ac magnetic field, select clusters with excitations resonant with the pump frequency can largely decouple from the remaining clusters and can be (almost) fully-described by a finite number of local spin operators. We can describe these resonantly-driven clusters with the MBL phenomenology, where an 'l-bit' (localized-bit) corresponding to the two-level cluster is comprised of a finite number of 'p-bits' (physical-bits) corresponding to the physical spins.[58] A model effective Hamiltonian for understanding the behavior of these bits is the disordered Ising spin chain[8]:

$$\mathcal{H} = -\sum_{i=1}^{L} h_i \sigma_i^x - \sum_{i=1}^{L-1} (J_i \sigma_i^z \sigma_{i+1}^z + \lambda_i \sigma_i^x \sigma_{i+1}^x) \qquad (6)$$

where the Ising and transverse interaction terms arise from the dipole-dipole coupling in the microscopic Hamiltonian. While the underlying experimental system is 3-dimensional, near the threshold of percolation among the clusters we can consider the bits as approaching the 1-d limit described by (6). The appropriateness of a strongly localized picture for the excitations is supported by the fact that the resonances are remarkably narrow ($Q = \frac{f_{\text{pump}}}{\Gamma} = \frac{202\,\text{Hz}}{2\,\text{mHz}} \approx 10^5$), notwithstanding the long range and large strength of the dipolar interaction.

The rotating reference frame of the strong pump field allows us to consider the system in a Floquet picture,[59] where the l-bits effectively form a network of co-rotating spins while the off-resonant clusters are far more weakly driven and form the continuum of excitations required for the appearance of Fano resonances. By measuring the structure of the resonances, we probe the dynamics of the localized cluster excitations, and we have shown that the dynamics are tunable, both by varying the strength of the strong pump field, and by introducing quantum mixing through a static transverse magnetic field. It has been suggested[11,60] that MBL would be observable in such systems via detailed measurements of the energy spectra of spatially localized operators. In our experiment, the requisite localized (spin) operators are localized in frequency, which is likely to be equivalent to spatial localization given the broad magnetic response spectrum of our disordered magnet as well as the very sharp holes burnt by the non-linear ac drive. Recent experiments in cold-atom systems[13,14] have addressed similar questions in lower dimensions, employing the degree of disorder rather than the strength of external longitudinal ac and transverse dc fields as the principal tuning parameters. For $LiHo_xY_{1-x}F_4$, residual thermal conductivity to outside heat reservoirs and the fundamentally three-dimensional nature of the material could, in the eventual long-time limit, result in thermalization and hence destruction of the MBL state. However, the measured behavior was stable on the scale of days, indicating that the timescale for destruction of the state is considerably longer. It is important to reiterate that these behaviors only appear when the crystal is weakly coupled to an external heat bath, and that energy transfer from the spins to phonons and hence the external bath is correspondingly slow. Increasing the efficiency of heat transfer destroys not only the localized state but also the underlying resonant cluster response (Fig. 1d).

It is important to consider the origin of the zero crossings of the Fano $q$ – a key result of our experiments - in terms of the ingredients of $q$, which are matrix elements linking the ground and excited states of the resonantly driven spin clusters to each other as well as to the bath formed by other clusters. The measurements characterize the spectral holes inserted by a nonlinear drive field into the continuum of magnetic excitations in a dense set of interacting dipoles. The large but finite lifetime of the excitations is due to a slight mixing between these (almost perfectly) localized excitations and the continuum formed by their ensemble. The mixing manifests itself in both the decay rate $\Gamma$ and the Fano $q$ parameter (c.f. Eq. (5)). The former is given by Fermi's golden rule:

$$\hbar\Gamma = 2\pi \sum_k V_k^2 \delta(\hbar\omega - E_k) \quad . \tag{7}$$

The q parameter is due to interference between processes taking the ground state of a localized subsystem to an excited state either directly or via another (nearly) localized subsystem[61]:

$$q = \frac{M_{g\alpha} + \mathbb{P}\sum_k \frac{M_{gk}V_k}{(\hbar\omega - E_k)}}{\pi \sum_k M_{gk}V_k \delta(\hbar\omega - E_k)} \quad , \tag{8}$$

where $M_{g\alpha} = \langle g|M|\alpha\rangle$ and $M_{gk} = \langle g|M|k\rangle$ are the matrix elements connecting the ground state to the discrete excited state and to the continuum, respectively. In (7) and (8), $V_k$ is the matrix element connecting the discrete excited state to the k$^{th}$ continuum state, which has energy $E_k$.[61] Inspection of Eq. 8 allows us to start to understand the zero crossing of $q$. In the nonlinear regime, the numerator has a modified matrix element $M_{g'\alpha'}$ describing how the longitudinal magnetic field couples the ground and excited states $g'$ and $\alpha'$ as modified by the drive field from $g$ and $\alpha$, while the denominator contains the product of the analogous matrix element $M_{gk}$ for off-resonant pairs of ground and excited states and the hopping terms $V_k$ between the resonant and off-resonant excited states. It is unlikely that the off-resonant $M_{gk}$ will be much changed by external ac and dc fields, and so – assuming that the principal value term in the numerator cancels to zero – sign changes in $q$ follow from a sign change either in $M_{g'\alpha'}$ or in $V_k$. A sign change in the latter actually would imply a zero in the former as well because without such a zero, $q$ would diverge at the critical pump (Fig. 5) or transverse (Fig. 6) field $h_c$. Where all terms in Eq. 8 are analytic near the zero crossing, and with the knowledge that $q$ is linear in $h_{pump}$ and $H_t$ near the zero crossing, we can draw a sharper conclusion, namely that $M_{g\alpha}$ scales like $(h-h_c)^{n+1}$ if $V_k$ scales like $(h-h_c)^n$. To determine $n$, we can invoke Eq. (7) and the experimental data which show no significant evolution in $\Gamma$ as a function of the parameters $h$ and $H_t$, leading to the conclusion that near the zero crossing of $q$, the exponent characterizing the bath coupling $V_k$ as a function of $h$ and $H_t$ is $n=0$. Therefore, what is driving the zero crossing is only a zero crossing of the matrix element $M_{g\alpha}$ connecting the ground and excited states of the localized subsystems. This means that as $h_{pump}$ crosses $h_c$, the incremental magnetization $\delta M$ due to mixing of ground $g(h)$ and excited $\alpha(h)$ states moves from in-phase to out-of-phase with small additional drive fields $\delta h$. It is reasonable to believe that where this occurs, the incremental magnetization due to the changing occupancies of the ground and excited states will be highest so that as we observe in the experiment, the zero crossing of $q$ will coincide with the maximum of the total susceptibility which sums diagonal (state occupancy-dominated) and off-diagonal ($M_{g\alpha}$-dominated) contributions. A further consequence of such considerations is that as the off-diagonal matrix element $M_{g\alpha}$ which accounts for the Fano effect grows from zero, it will also account for an ever larger fraction of the dissipation measured directly at the pump frequency. This allows the proposal of a phenomenological form (solid blue line) for the phase angles plotted in Figs. 5b and 6b:

$$\phi'(q) = \tan^{-1}\left(\frac{q^2}{q^2+c^2+(h_{\text{pump}}/d)^2} \cdot \frac{\chi''_{\text{drive}}(f=f_{pump})}{\chi'_{\text{drive}}(f=f_{pump})}\right) \quad (9)$$

where $c=0.12$ represents the intrinsic contributions of the off-diagonal matrix element and $d=1.4$ Oe incorporates the thermal effects associated with changes in pump amplitude. We note that Eq. (9) explicitly incorporates the experimental observation that the vanishing of the dissipation (and hence of φ) coincides with the zero-crossing of $q$.

The field scale $H_c$ can be connected to the microscopic properties of Li(Ho,Y)F$_4$ by comparison with an exact diagonalization of the full Hamiltonian for a pair of coupled Ho$^{3+}$ atoms[37], where there is a crossing of the lowest coupled electro-nuclear levels for nearest-neighbor spins positioned in the ab plane at (100) or (010) relative to each other. The effects of these pairwise level crossings can be seen in the measured linear longitudinal susceptibility $\chi_{zz}(H_t)$[24,37]. The characteristic dynamics of the pairwise susceptibility shift as a function of classical (temperature) and quantum-mechanical (transverse field) energy scales, moving from 0.9 T at 70 mK to 0.5 T at 150 mK. We expect the same level crossings to impact the dynamics of the larger spin clusters addressed by the pump-probe measurements described here, with the transverse field scales reported here reduced in comparison to those obtained in the pairwise calculation due to additional interactions from further-neighbor spins.

Given that we have not performed similar experiments for compositions x other than $x$=4.5%, we only can speculate as to the possible evolution of the localized cluster behavior as a function of the holmium ion concentration. As $x$ is increased to 16.7%, only spin-glass-like behavior has been seen[40,43,62], suggesting that placing the system in the "antiglass" state, which is a prerequisite for hole burning, would be difficult. By $x$=20%, a combination of quantum effects and random-field physics for a more nearly ferromagnetic system changes the nonlinear response[63], and isolated clusters even in the limit of thermal isolation are unlikely, and become impossible at the onset of long-range ferromagnetic order at $x$~30%.[43]

### Conclusions

Via hole burning, we have demonstrated localization of excitations among interacting magnetic dipoles. For strong coupling to the heat bath, previous work has shown a trend towards spin freezing at low temperatures[25,49], and the current work (Fig. 1d) shows no spectral hole burning in this strong coupling limit, corresponding to delocalized excitations. Reducing the coupling to the external thermal bath induces a transition to a spin liquid state[25,51] where the excitations are localized to a very high degree, with $Q$~$10^5$ for the spectral hole. The small mixing between excited states of different spin clusters leads to a Fano effect, whose phenomenology for this complex system is remarkably simple, with a vanishing $q$ coinciding with an inflection point in the ac magnetization induced by the drive field responsible for the hole-burning. The experiments show the value of hole-burning for characterizing many-body localization[4,7,8], and also demonstrate how nonlinear quantum dynamics can reveal emergent two-level systems.

## Methods

We cooled a single crystal of LiHo$_{0.045}$Y$_{0.955}$F$_4$ in a helium dilution refrigerator and measured its ac magnetic susceptibility for frequencies from 1 to 2000 Hz. The thermal coupling between the crystal and the heat reservoir – the mixing chamber of the dilution refrigerator – was in the "weakly coupled" regime described by Ref. 25, thereby maximizing quantum fluctuations. We employed a pump-probe technique[24,25,31] to access the nonlinear response regime. In this configuration, a two-frequency ac magnetic field is applied along the Ising axis of the crystal. A strong (0.2-0.6 Oe) field $h_{\text{pump}}\cos(2\pi f_{\text{pump}}t)$ excites clusters at the pump frequency $f_{\text{pump}}$. Simultaneously, a 20 mOe probe field is swept through a range of frequencies to yield a linear response from the crystal. The net response of the crystal is then sensed by an inductive pickup coil. Due to the extremely narrow separation between the pump and probe frequencies (as small as 1 mHz), disentangling the responses required the development of a 2-stage lock-in technique, where the combined response is passed into a commercial lock-in amplifier tuned to the probe frequency, and the resulting output is sampled by a computer and passed through a software lock-in detector tuned to $\Delta f = f_{\text{pump}} - f_{\text{probe}}$. It should be noted that the sampling step requires a minimum of $1/\Delta f$ seconds; all of the data reported here were sampled for a minimum of $2/\Delta f$ seconds.

## Acknowledgments

We thank G. Refael, Michael Buchhold, and Markus Müller for helpful discussions. The work at Caltech was supported by US Department of Energy Basic Energy Sciences Award DE-SC0014866.

**Author Contributions**

D.M.S. and T.F.R. performed the measurements. All authors participated in the analysis of the data and the writing of the manuscript.

**Additional information**

Correspondence and requests for materials should be addressed to T.F.R., tfr@caltech.edu.

**Competing financial interests**

The authors declare no competing financial or non-financial interests.

**Figure Captions**

1. Schematic of spin configuration and transition pathways in pumped LiHo$_{0.045}$Y$_{0.955}$F$_4$. **a** Tightly-bound spin clusters embedded in a dilute spin bath. Closely-separated spins form a tightly bound cluster which can be magnetized in a nonlinear response regime by a strong ac pump field. Spatially adjacent clusters can interact directly. **b** Schematic of cluster excitation pathways. Clusters of different sizes have different energy gaps, so choice of pump frequency acts as a size selector. Direct interactions between excited clusters and off-resonance clusters give rise to Fano resonance behavior. **c** Magnetization amplitude (top) and phase angle (bottom) for three different excitation frequencies. **d** Pump-probe measurements in two thermodynamic regimes. When LiHo$_{0.045}$Y$_{0.955}$F$_4$ is weakly coupled to an external heat reservoir, the low temperature state is dominated by the quantum cluster response, giving rise to strong Fano resonances. When it is strongly coupled to an external heat reservoir, it exhibits extended glassy behavior including the absence of hole-burning and Fano resonance effects[25]. Both traces were measured at $T$=100 mK, $f_{pump}$=20 Hz, and $h_{pump}$=0.5 Oe.

2. Linear absorption spectra of LiHo$_{0.045}$Y$_{0.955}$F$_4$ as a function of temperature; excitation field has 20 mOe amplitude. **a** Measured absorption in the presence of a 0.3 Oe pump field at $f_{pump}$ = 202 Hz and zero transverse field. Curves are fits to Fano resonance forms, Eq. 5 in the text. The points at $f$=$f_{pump}$ (open symbols) are omitted from the fits (see text for details). **b** Absorption normalized with respect to the response at $\Delta f = 30$ mHz as a function of frequency and temperature. **c** Linewidth of the resonance, as determined from fitting to the Fano model, vs. temperature $T$. Line is a fit to an intrinsic linewidth of 1.7 mHz plus exponential thermal broadening with $\Delta = 740$ mK. **d** Fano parameter $q$ vs.

temperature, showing the suppression of coupling to the bath at the lowest $T$. Lines are guides to the eye.

3. Linear absorption as a function of pump amplitude at $T = 0.11$ K for a 202 Hz pump. **a,b** Measured imaginary and real susceptibilities (points), and fits to a Fano resonance form as a function of pump amplitude at zero transverse field. Increasing the pump amplitude tunes the resonant behavior, at the cost of increased decoherence. **c** Normalized absorption as a function of frequency and pump amplitude.

4. Linear absorption as a function of transverse field at $T = 0.11$ K for a 202 Hz pump. **a,b** Measured imaginary and real susceptibilities (points) and fits to a Fano resonance as a function of transverse field for a fixed 0.3 Oe pump. Transverse field-induced quantum tunneling tunes the resonant behavior without a corresponding increase in decoherence. **c** Normalized absorption as a function of frequency and transverse field.

5. Evolution of resonant behavior as a function of pump amplitude at $T = 0.11$ K and $H_t=0$ for a 202 Hz pump. **a** Fano parameter $q$ vs. drive amplitude showing a continuous evolution including a smooth crossing through zero. Dashed line is a guide to the eye. **b** Evolution of the phase of the complex susceptibility at $f = f_{pump}$ for the nonlinear (open symbols) and linear (filled symbols) responses. The zero crossings of $q$ are associated with a local minimum in the dissipation, and a corresponding minimum in the phase shift of the linear probe response as the probe frequency approaches the pump frequency. $\phi(q) = \tan^{-1}\frac{q^2}{1-q^2}$ (blue dotted curve) follows from Eq. 5 in the text, while the solid blue curve (see Eq. 8 and associated discussion in text) provides a better description of the experimental data. **c** Real susceptibility $\chi'$ measured directly at $f=f_{pump}$ (open symbols) and determined by extrapolating the fitted Fano resonance to $f=f_{pump}$ (filled symbols), showing the contrast in behavior between the nonlinear and linear responses respectively. **d** Fano linewidth $\Gamma$ vs. ac drive $h_{pump}$. Increasing the drive amplitude broadens the linewidth and hence reduces the oscillator $Q$.

6. Evolution of resonant behavior as a function of transverse field at $T = 0.11$ K for a 202 Hz/0.3 Oe pump. **a** Fano parameter $q$ vs. transverse field showing a continuous evolution including a smooth crossing through zero. Dashed line is a guide to the eye. **b** Evolution of the phase of the complex susceptibility at $f = f_{pump}$ for the nonlinear (open symbols) and linear (filled symbols) responses. As with the $h_{pump}$ dependence shown in Fig. 5, the zero-crossing of $q$ is associated with a vanishing of the dissipation in the linear response with the same functional form, demonstrating universal behavior from two disparate tuning parameters. **c** Real susceptibility $\chi'$ measured directly at $f=f_{pump}$ (open symbols) and determined by extrapolating the fitted Fano resonance to $f=f_{pump}$ (filled symbols), showing a small but apparent distinction in the evolution of the nonlinear and linear responses. **d** Fano linewidth $\Gamma$ vs. transverse field. In contrast to the behavior as a function of $h_{pump}$, increasing $H_t$ does not change the linewidth.

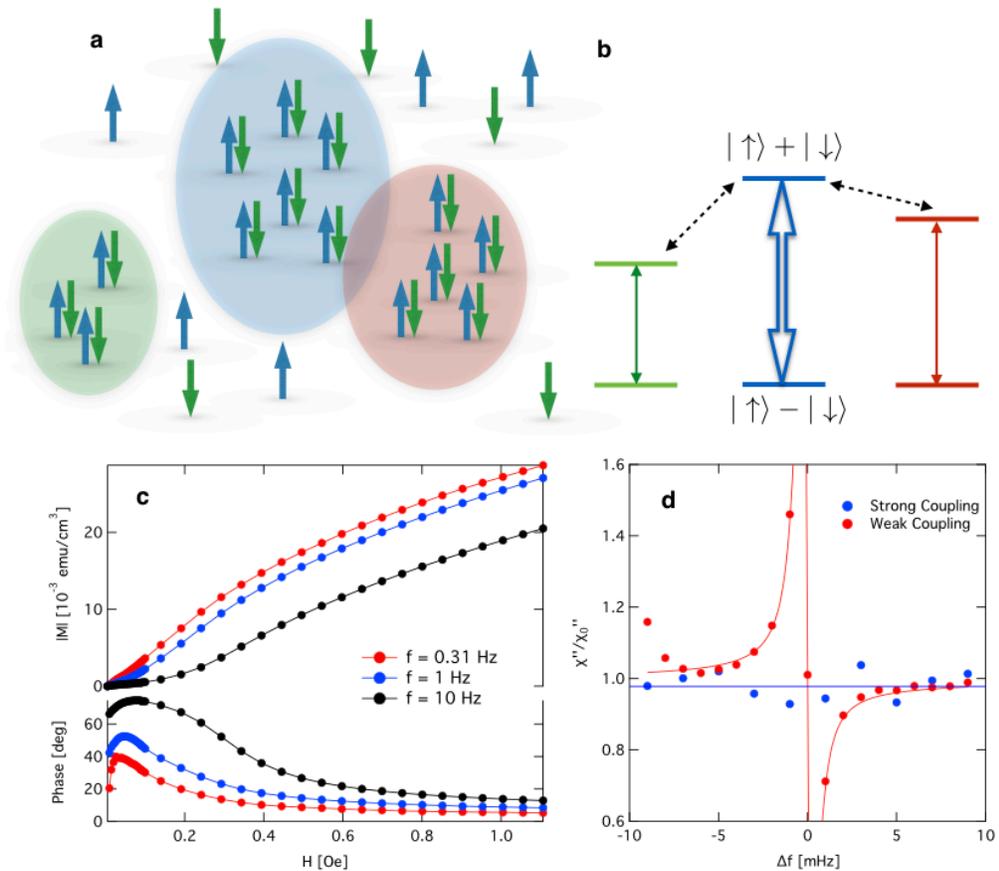

**Figure 1**

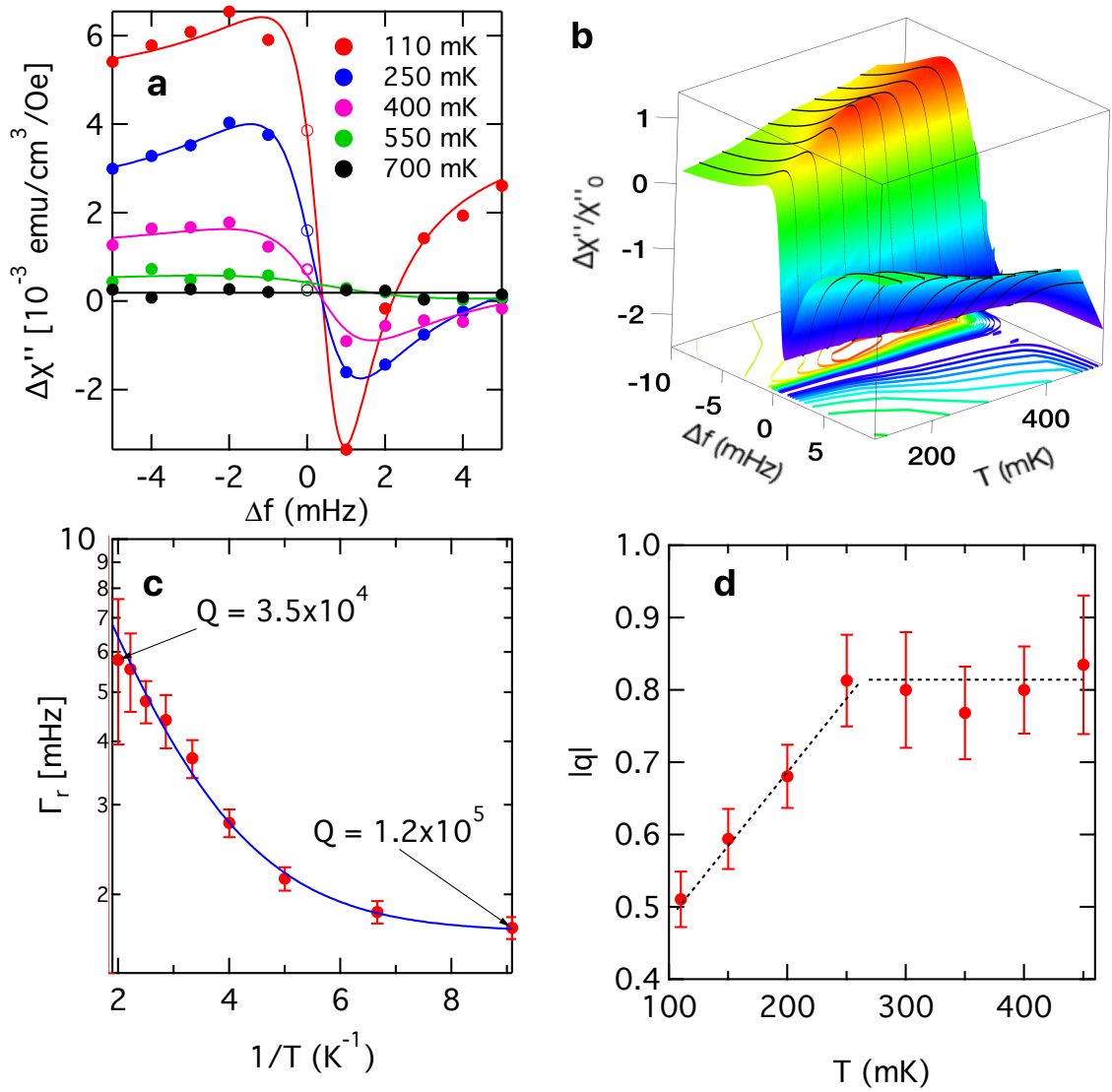

**Figure 2**

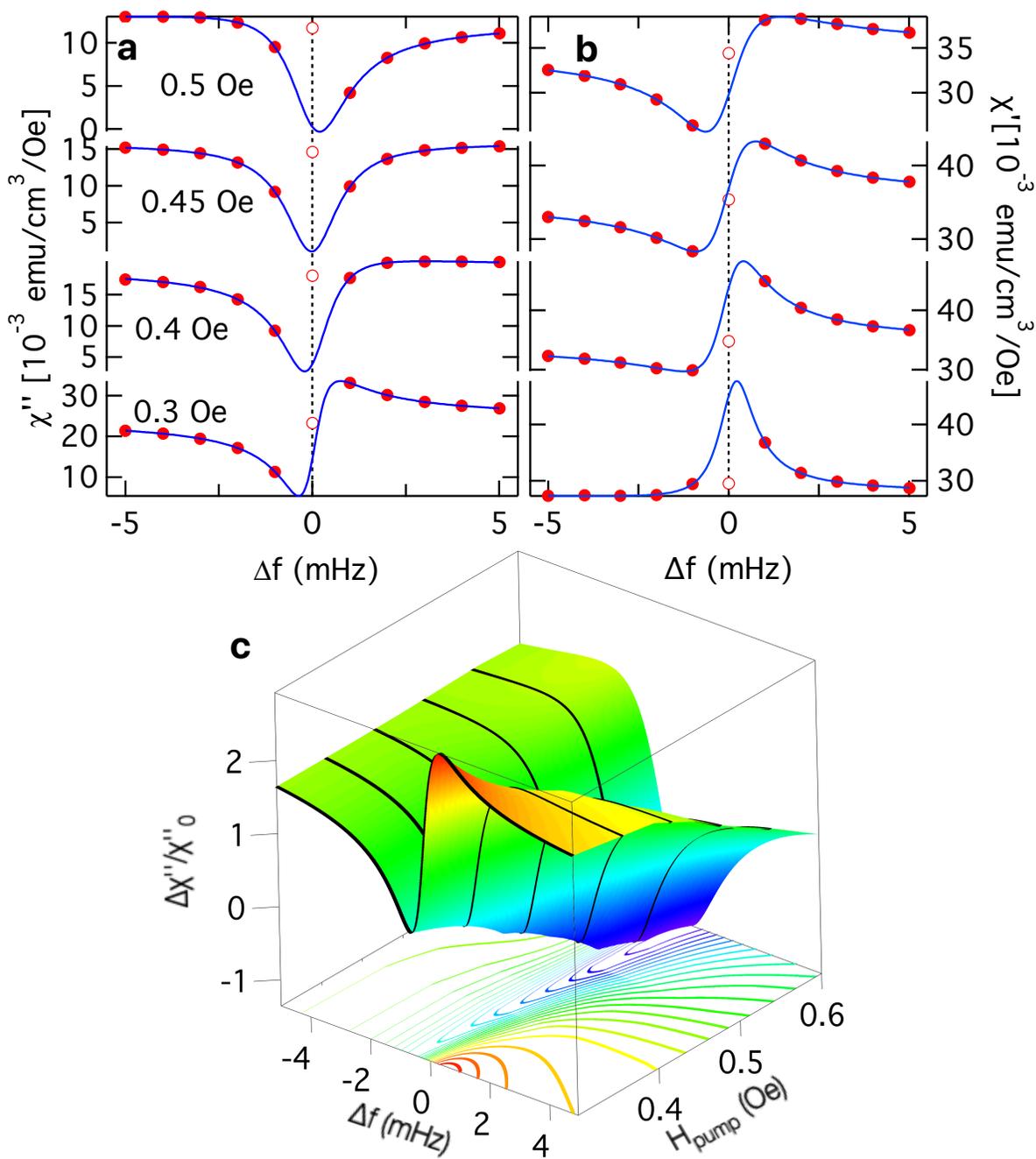

**Figure 3**

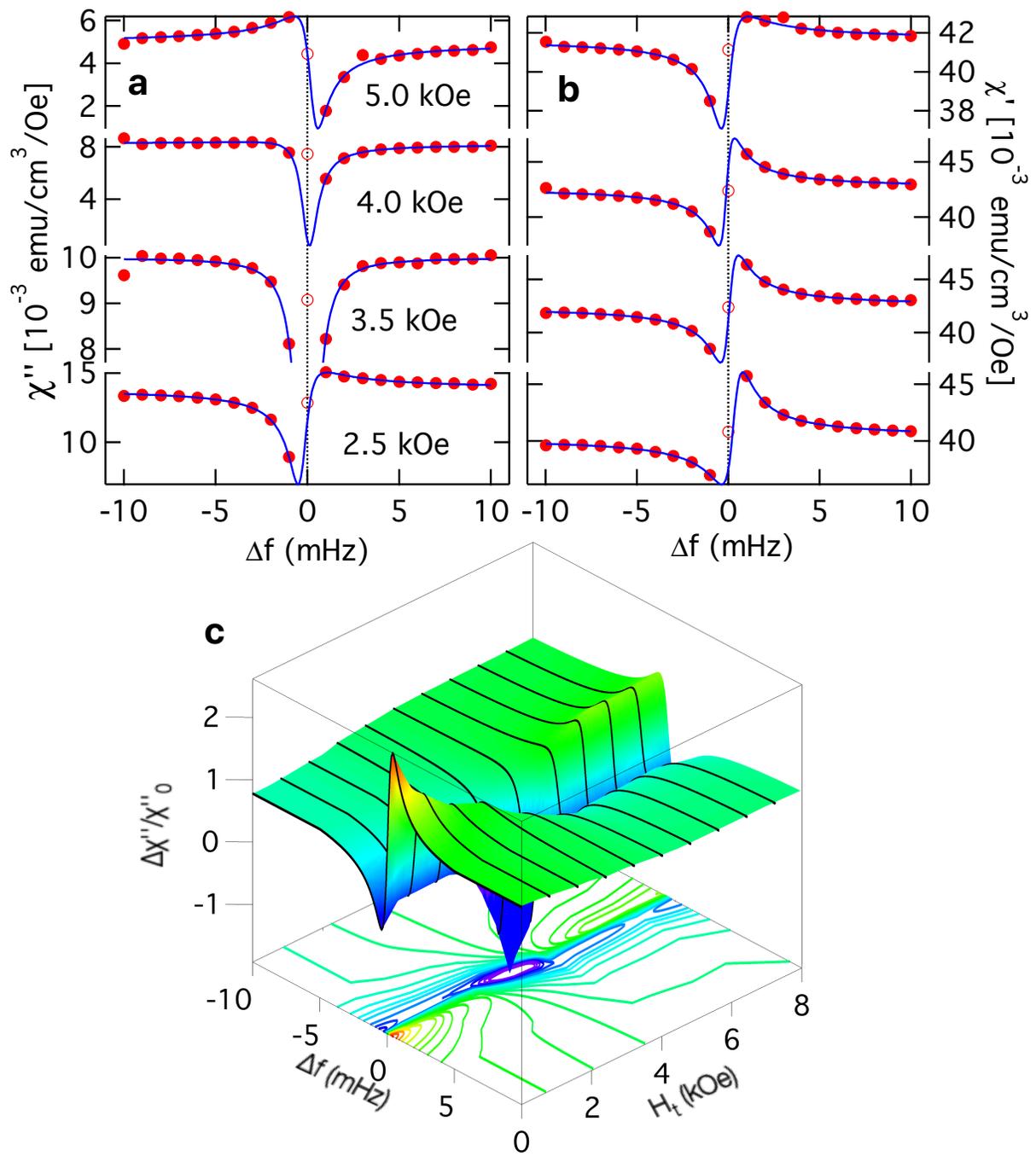

**Figure 4**

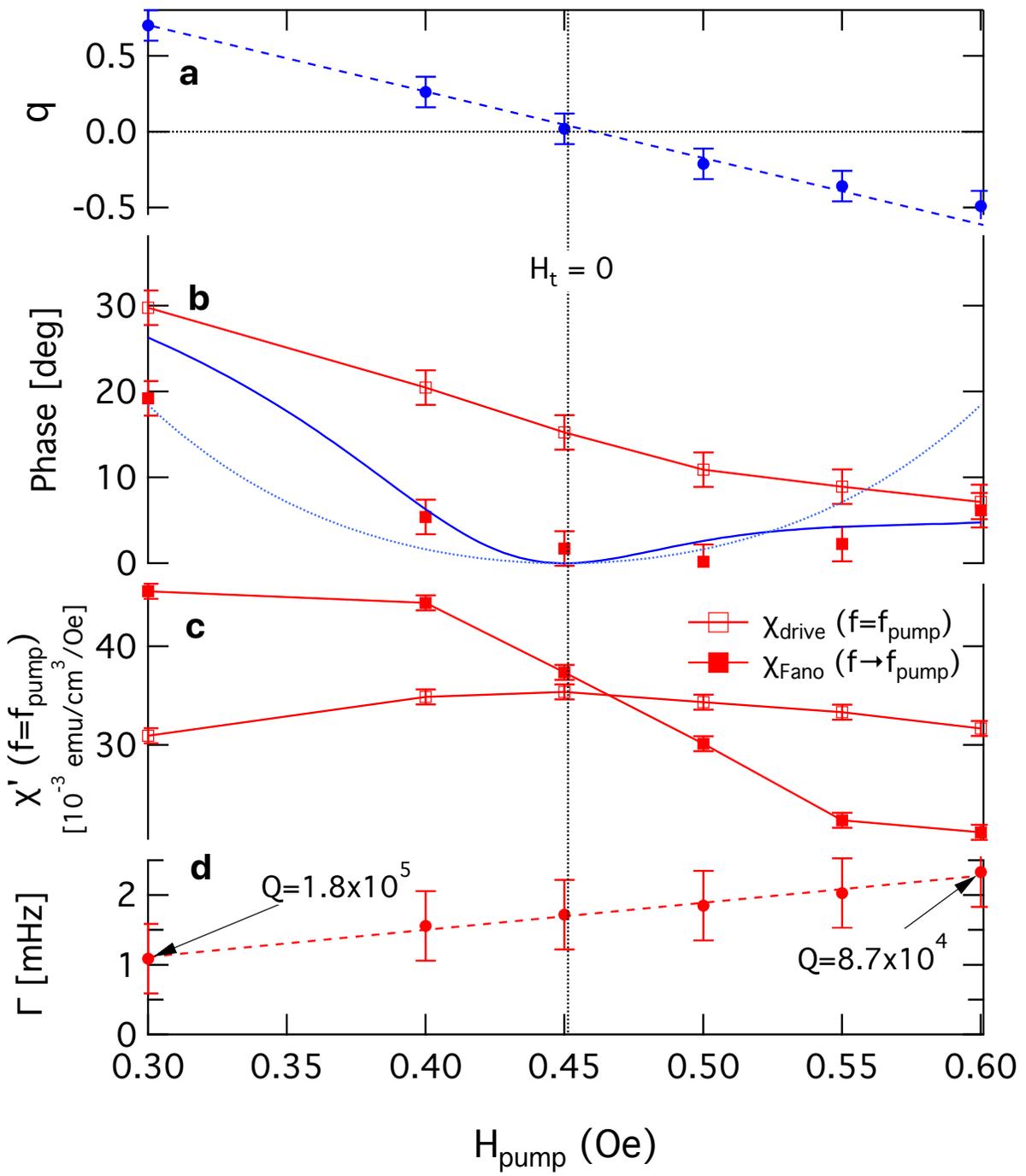

**Figure 5**

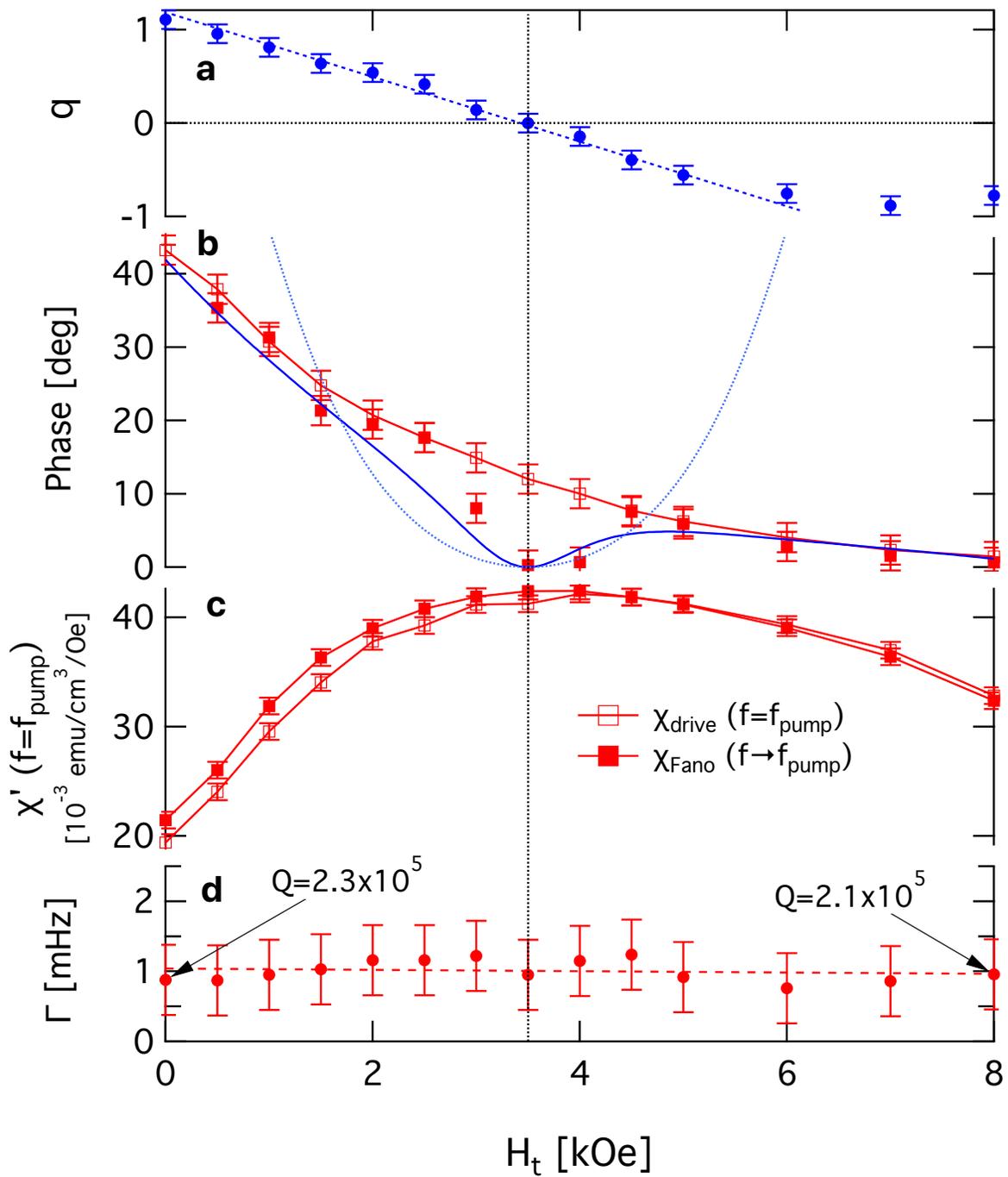

**Figure 6**